\documentclass[twocolumn,showpacs,preprintnumbers]{revtex4-2}
\usepackage[caption=false]{subfig}
\usepackage{amsmath}  
\usepackage{amsfonts}  
\usepackage{graphicx} 
\usepackage{enumerate}
\usepackage{tikz}
\usetikzlibrary{shapes,arrows, arrows.meta, calc, positioning}
\usepackage[colorlinks=true,linkcolor=blue,urlcolor=blue,citecolor=blue]{hyperref}
\begin{document}
\newcommand{\ds}{\displaystyle}
\newcommand{\be}{\begin{equation}}
\newcommand{\en}{\end{equation}}
\newcommand{\bea}{\begin{eqnarray}}
\newcommand{\ena}{\end{eqnarray}}

\title{Revisiting Wormhole Solutions in Unimodular Gravity: Energy Conditions and Exotic Matter Requirements}
\author{Mauricio Cataldo}
\altaffiliation{mcataldo@ubiobio.cl} 
\affiliation{Departamento de
F\'\i sica, Universidad del Bío-B\'io, Casilla 5-C,
Concepci\'on, Chile.\\}
\author{Norman Cruz}
\altaffiliation{norman.cruz@usach.cl} 
\affiliation{Departamento de Física, Universidad de Santiago de Chile, Avenida Ecuador 3493, Santiago, Chile, \\}
\affiliation{Center for Interdisciplinary Research in Astrophysics and Space Exploration (CIRAS), Universidad de Santiago de Chile, Av. Libertador Bernardo O’Higgins 3363, Estación Central, Chile. \\}

\date{\today} 
\begin{abstract}
The paper entitled ``Unimodular Gravity Traversable Wormholes'' by Agrawal et al. examined the properties of barotropic wormholes without tidal forces within the framework of Unimodular Gravity. This comment demonstrates that their conclusion regarding the possibility of sustaining such wormhole configurations with ordinary matter is not entirely accurate. We establish that exotic matter remains necessary for these wormhole solutions in Unimodular Gravity, in accordance with the long-established theoretical constraints already identified in General Relativity.

\vspace{0.5cm}
\end{abstract}
\smallskip
\maketitle 
\preprint{APS/123-QED}

The recent work by Agrawal et al.~\cite{Agrawal} presents an advancement in the field of wormholes deriving, for the first time, traversable wormhole solutions within the framework of Unimodular Gravity. Their approach examines wormholes free of tidal forces through two key assumptions: a constant redshift function and barotropic equations of state for the pressures of the supporting anisotropic fluid. These assumptions simplify the field equations while preserving the asymptotically flat limit condition. The analysis focuses on verifying the energy and traversability conditions within the framework of Unimodular Gravity, concluding that these solutions satisfy all metric conditions of traversability, as well as the classical energy conditions, thus observing that these wormholes succeed in evading the need for exotic matter.

In this Comment, we demonstrate that, contrary to these claims, wormhole configurations without tidal forces in Unimodular Gravity necessarily require exotic matter. To establish this result, we shall rederive the complete wormhole solution presented in their work, and subsequently examine the precise conditions governing the energy requirements that must be satisfied by these spacetime configurations.

The field equations of Unimodular Gravity are considered in the form~\cite{Agrawal}
\begin{equation}
R_{\mu\nu} - \frac{R}{4} g_{\mu\nu} = \kappa \left(T_{\mu\nu} - \frac{T g_{\mu\nu}}{4}\right), \label{eccu}
\end{equation}
where $\kappa=8 \pi G$, $R_{\mu \nu}$ is the Ricci tensor, $T_{\mu \nu}$ is the energy-momentum tensor, and $T$ denotes its trace.

The standard formulation for wormholes without tidal forces is characterized by the spacetime~\cite{Morris}:
\begin{equation}
ds^2 = dt^2 - \frac{dr^2}{1 - \frac{b(r)}{r}} - r^2 \left(d\theta^2 + \sin^2\theta d\phi^2 \right),\label{whm}
\end{equation}
where $b(r)$ is the shape function. For this metric to describe a traversable wormhole, it must satisfy the following conditions at the throat $r_0$: $b(r_0) = r_0$,
$b'(r_0) \leq 1$, $b'(r) < \frac{b(r)}{r}$ and $b(r) < r$, where the prime represents the derivative with respect to the radial coordinate $r$.

The field equations~(\ref{eccu}) for the metric~(\ref{whm}) are given by:
\begin{align}
\frac{2b'}{\kappa r^2} &=   3\rho + p_r + 2p_t, \label{tt} \\
\frac{2b}{\kappa r^3} &=   \rho - p_r + 2p_t, \label{rr}\\
-\frac{2(2b - rb)'}{\kappa r^3}  &= \rho + 3p_r - 2p_,.  \label{pp}
\end{align}   
where $\rho$, $p_r$ and $p_t$ represent the energy density, radial pressure, and tangential pressure of the anisotropic fluid supporting the wormhole.

From Eqs.~(\ref{rr}) and~(\ref{pp}), we derive expressions for the pressures in terms of the shape function and energy density:
\begin{eqnarray}
\kappa p_r = \frac{-\kappa \rho r^3 + b' r - b}{ r^3}, \label{ec1}\\
\kappa p_{t}= \frac{-2\kappa  \rho r^3 + b' r + b}{2 r^3}. \label{ec2}
\end{eqnarray}
Upon substituting these two expressions into Eq.~(\ref{tt}), we demonstrate that the equation is identically satisfied, revealing that the system of unimodular gravitational field equations~(\ref{tt})-(\ref{pp}) contains only two independent equations, while involving four unknown functions: $b(r)$, $\rho(r)$, $p_r(r)$, and $p_t(r)$. To solve this underdetermined system, additional constraints must be imposed. Following standard practice, the authors consider barotropic equations of state $p_r = \alpha\rho$ and
$p_t = \beta p_r$ for the pressures, where $\alpha$ and $\beta$ are constants.

Based on these assumptions, the shape function, the energy density and the pressures are given by
\begin{eqnarray}
b(r) &=& Cr^{\frac{\alpha(2\beta+1)+3}{\alpha(2\beta-1)+1}}, \\
\kappa \rho(r) &=& \frac{2 C r^{-\frac{4\alpha(\beta-1)}{1+(2\beta-1)\alpha}}}{1 + \alpha(2\beta - 1)}, \label{ec rho}\\
\kappa p_r(r) &=& \frac{2 \alpha C r^{-\frac{4\alpha(\beta-1)}{1+(2\beta-1)\alpha}}}{1 + \alpha(2\beta - 1)}, \label{ec pr} \\
\kappa p_t(r) &=& \frac{2 \alpha \beta C r^{-\frac{4\alpha(\beta-1)}{1+(2\beta-1)\alpha}}}{1 + \alpha(2\beta - 1)}, \label{ec pt}
\end{eqnarray}
where $C$ is an integration constant. Putting $C = r_{0}^{-\frac{2(\alpha + 1)}{1 + \alpha(2\beta - 1)}}$, we have $b(r_0)=r_0$ and Eq.~(\ref{whm}) takes the form
\begin{equation*}
ds^2 = dt^2 - \frac{dr^2}{1 -\left(\frac{r}{r_{0}}\right)^{\frac{2(\alpha + 1)}{1 + \alpha(2\beta - 1)}}} - r^2 \left(d \theta^2 + \sin^2 \theta d \phi^2 \right).
\end{equation*}
Note that, for this metric to represent a wormhole, the following condition must be satisfied
\begin{equation}
\frac{\alpha + 1}{1+\alpha(2\beta -1)} < 0, \label{ca}
\end{equation}
so the parameters satisfy the following inequalities:
\begin{eqnarray}
\alpha < -1, \beta < \frac{\alpha - 1}{2\alpha},\\ 
-1 < \alpha, \alpha < 0, \frac{\alpha - 1}{2\alpha} < \beta, \\
0 < \alpha, \beta < \frac{\alpha - 1}{2\alpha}.
\end{eqnarray}
To contrast this solution with the energy conditions, we will set $C=1$, which implies that the wormhole throat is located at $r_0=1$.

To verify fulfillment of the energy conditions, we must evaluate whether the anisotropic matter distribution sustaining the wormhole satisfies~\cite{Visser}:
\begin{enumerate}
\item Null Energy Condition (NEC): \\ $\rho + p_r \geq 0$ and $\rho + p_t \geq 0$.
\item Weak Energy Condition (WEC): \\ $\rho \geq 0$, $\rho + p_r \geq 0$ and $\rho + p_t \geq 0$.
\item Dominant Energy Condition (DEC): \\ $\rho \geq |p_r|$ and $\rho \geq |p_t|$.
\item Strong Energy Condition (SEC):\\ $\rho + p_r + 2p_t \geq 0$.
\end{enumerate}
In other words, all these conditions must be satisfied simultaneously for any values of the parameters $\alpha$ and $\beta$. Since we have analytical expressions for this solution, we can use them to concretely write the inequalities that define the energy conditions. Thus, from expressions~(\ref{ec rho})-(\ref{ec pt}), we obtain:
\begin{eqnarray}
\rho + p_r = \frac{2 r^{-\frac{4\alpha(\beta-1)}{2\alpha\beta-\alpha+1}} (\alpha+1)}{\kappa(1 + \alpha(2\beta - 1)}, \label{nec1}\\
\rho + p_l = \frac{2 r^{-\frac{4\alpha(\beta-1)}{2\alpha\beta-\alpha+1}}  (\alpha\beta+1)}{\kappa(1 + \alpha(2\beta - 1)}. \label{nec2}
\end{eqnarray}
It is evident from Eq.~(\ref{nec1}) that NEC and WEC are not satisfied, since condition~(\ref{ca}) implies that $\rho + p_r<0$. The violations of NEC and WEC are inherent to the mathematical structure of the studied solution, since for any combination of parameters $\alpha$ and $\beta$ that produces a valid wormhole, NEC and WEC are not fulfilled.

In conclusion, despite the claim that these solutions satisfy all energy conditions, we have rigorously demonstrated that even in Unimodular Gravity, barotropic wormholes without tidal forces necessarily cannot be supported by ordinary anisotropic matter.

\begin{acknowledgments}
M.C. acknowledges the support from the Dirección de Investigación y Creación Artística at the Universidad del Bío-Bío through grants No. RE2320220 and GI2310339. N.C. acknowledges the support from ANID-CHILE through Fondecyt grant N°1250969.
\end{acknowledgments}

\end{document}